\documentclass[aps,notitlepage]{revtex4-1}
\usepackage{graphicx}
\usepackage{amsmath}
\usepackage[colorlinks]{hyperref}
\usepackage[caption=false]{subfig}
\usepackage{setspace}

\begin{document}

\title{CHARGED HIGGS PAIR PRODUCTION IN THDM THROUGH PHOTON-PHOTON COLLISIONS AT THE ILC}

\author{Nasuf SONMEZ}
\affiliation{Ege University, Izmir, Turkey}


\date{\today}

\begin{abstract}
In this study, the charged Higgs pair production is analyzed for the minimal extension of the standard model called two-higgs-doublet model.
The process $\gamma\gamma\rightarrow H^+H^-$ is calculated at the tree level for the ILC and the numerical analysis is presented for various parameters.
The production rate of the charged Higgs boson pair as a function of center-of-mass (CM) energy and the differential cross section as a function of angle between photon and positive charged Higgs boson is presented.
The cross section gets high at the low charged Higgs mass and low CM energies.
The total integrated cross section of the process is also calculated at a $e^+e^-$-collider by convoluting the $\gamma\gamma\rightarrow H^+H^-$ subprocess with the photon luminosity of the backscattered photons.
The total integrated cross section peaks around $\sqrt{s}=650 \;\text{GeV}$ and have a value of $1.4 \;\text{pb}$ for $m_{H^\pm}=100\;\text{GeV/c}^2$.
Charged Higgs detection is very important sign for the new physics and the results shows the potential of the ILC for the search of the new physics signals.
\end{abstract}


\maketitle


\setstretch{1.5}

\section{Introduction}
Over the last couple of decades, many extensions to cure the quadratic divergence at the scalar sector in the Standard Model (SM) have been proposed and the implications of the new physics have been studied intensively.
One possible extension of the SM is to add a second Higgs doublet to the scalar sector.
The new scalar doublet has the same quantum number with the Higgs doublet and they together give mass to leptons, quarks and electroweak bosons.
This model is called Two Higgs Doublet Model (THDM).
In a general THDM, there will be two charged Higgs bosons $(H^\pm)$ and three neutral Higgs bosons $(h^0,A^0,H^0)$  \cite{Branco:2011iw,Gunion:1989we}.
Where $h^0$ is a SM like particle and its properties bear resemblance to the discovered Higgs boson.
In the model, $h^0$ along with $H^0$ are CP-even, whereas $A^0$ is CP-odd Higgs boson.

There has been a long time effort to observe a hint associated with a charged Higgs boson in the past and current experiments.
However, it is not yet discovered at the LEP, Tevatron and even at the LHC and the search is still going on. 
The LEP experiment excluded the charged Higgs boson with mass below $80 \;GeV$ (Type II scenario) or $72.5 \;GeV$ (Type I scenario, for pseudo-scalar masses above $12 \;GeV$) at the 95 \% confidence level.
If it is assumed that $BR(H^+\rightarrow \tau^+\nu)=1$, then charged Higgs mass bound increased to $94\;GeV$ for all $\tan\beta$ values. \cite{Abbiendi:2013hk}.
The Tevatron experiments D0 \cite{Abazov:2008rn,Abazov:2009ae,Abazov:2009wy} and CDF \cite{Aaltonen:2009ke} excluded charged Higgs mass in the range of $80\;GeV<m_{H^\pm}<155 \;GeV$ at the 95 \% confidence level, assuming $BR(H^+ \rightarrow c\bar{s})=1$.
Charged Higgs search is also studied at the LHC in the decay of top quark \cite{Aad:2012tj,Chatrchyan:2012vca} and upper limits are set for the $BR(t\rightarrow H^+ b)$ and $BR(H^+\rightarrow \tau\nu)$, respectively.

Nowadays, there is an ongoing effort for another project named International Linear Collider (ILC) where $e^+e^-$, $e^-e^-$ and $\gamma e$ collisions are studied.
The main task at the ILC is to complement the LHC results, and also search for clues in (BSM) such as supersymmetry, extension of the scalar sector and exotic models.
ILC is designed to study the properties of the new particles and the interactions they make according to the theory.
As it is expected, linear colliders compared to the LHC have cleaner background and it is possible to extract the new physics signals from the background more easily.
One option being considered for the ILC is to add $\gamma\gamma$-collider with the center of mass energy $\sqrt{s}=250-2000\;{GeV}$ with an integrated luminosity of the order of $100\;fb^{-1}$ yearly \cite{Behnke:2013xla, Behnke:2013lya}. 

A lot of effort have been invested in pair production of the charged Higgs boson at the hadron colliders, the production cross section reaches up to 40 femto-barn \cite{Jiang:1997cg, Hespel:2014sla}.
Therefore, the pair production of the charged Higgs boson is also possible at the linear colliders where the production rate is higher.
At a $\gamma\gamma$-collider, the production rate could be much higher than $e^+e^-$-collisions at the tree level.
In Ref. \cite{Hashemi:2013sja}, the process $e^+e^-\rightarrow H^+H^-$ for THDM is studied at the tree level, 
the $\gamma\gamma \rightarrow H^+H^-$ process have been studied before at the Born level and including only the Yukawa corrections \cite{Ma:1996nq}.
Discovery potential and detection signatures of the charged Higgs pair are studied in \cite{BowserChao:1993ji}.
Even though the process is analyzed including the Yukawa corrections, the born level process still have rich physics results and needs a detailed study in the light of recent constraints on the charged Higgs production.
In this work, the analytical and numerical calculations for the charged Higgs pair production are presented at the born level.
The total cross section as a function of the center-of-mass energy for the charged Higgs pair and the angular distribution of the cross section are calculated and analyzed for various parameters.
The $m_{H^\pm}$ dependence of the process is discussed and presented for two distinct energies.
In addition to these, the total cross section is calculated in a $e^+e^-$-collider for up to $\sqrt{s}=2\;{TeV}$ as a function of $\sqrt{s}$.

The content of this paper is organized as follows. 
In Sec. \ref{sec:2}, the scalar sector and the charged Higgs boson in THDM are discussed. 
In Sec. \ref{sec:3}, analytical expressions of the amplitude for the pair production of the charged Higgs and the kinematics of the scattering, the total cross section in $e^+e^-$ machine are given.
In Sec \ref{sec:4}, numerical results of the total cross section for the benchmark point are discussed.
The conclusion is drawn in Sec \ref{sec:5}.


\section{The scalar sector of the THDM}
\label{sec:2}

Adding another Higgs doublet to the scalar sector of the SM introduces in total 14 free parameters. 
In a general THDM, the parameters $m_{11}$, $m_{22}$ and $\lambda_{1,2,3,4}$ are real while $m_{12}$ and $\lambda_{5,6,7}$ are complex. 
We denote the Higgs doublets as 
\begin{equation}
\Phi_1 = 
    \begin{pmatrix}
   \phi_{1}^+  \\
	\phi_{1}^0\\ 
	    \end{pmatrix},    
    \;\;\;\;\;\;\;\;\;
    \Phi_2 = 
    \begin{pmatrix}
    \phi_{2}^+  \\
     \phi_{2}^0  \\
    \end{pmatrix},    
\end{equation}
and the vacuum expectation values are $<\Phi_1>=\nu_1/\sqrt{2}$ and $<\Phi_2>=\nu_2/\sqrt{2}$, respectively.
In the theory, both of the doublets have the same charge assignment, so that they could couple to leptons and quarks as in the SM.
The scalar potential of the THDM is given in Eq. \ref{eq:eq2}.
\begin{eqnarray}
V(\Phi_1,\Phi_2)=&&m_{11}^2 | \Phi_1|^2+m_{22}^2|\Phi_2|^2 - \left[ m_{12}^2   \Phi_1^{\dagger} \Phi_2 +h.c. \right] \nonumber \\ 
&+& \frac{\lambda_1}{2}| ( \Phi_1^{\dagger} \Phi_1 )^2  + \frac{\lambda_2}{2} ( \Phi_2^{\dagger} \Phi_2 )^2 +\lambda_3 | \Phi_1 |^2 | \Phi_2 |^2 +\lambda_4 | \Phi_1^{\dagger} \Phi_2 |^2 \label{eq:eq2} \\
&+& \left [  \frac{\lambda_5}{2} (\Phi_1^{\dagger} \Phi_2)^2 + \left ( \lambda_6 (\Phi_1^{\dagger} \Phi_1 ) + \lambda_7 (\Phi_2^{\dagger} \Phi_2 )  \right) \Phi_1^{\dagger} \Phi_2 + h.c.  \right] .\nonumber 
\end{eqnarray}

If it is assumed that the electromagnetic gauge symmetry is preserved in THDM, we could easily make a $SU(2)$ rotation on $\Phi_1$ and $\Phi_2$ doublets in such a way that the vev's of the two doublets are aligned in the $SU(2)$ space and $\nu = 246\;{GeV}$ will reside completely in the neutral component of one of the Higgs doublets \cite{Craig:2013hca}.
A special case appears in this alignment limit where $\Phi_1$ and $\Phi_2$ in the mass matrix go to zero and the lighter CP even Higgs boson $h^0$ is nearly indistinguishable from the Higgs boson of the Standard Model.

The CP violation and the flavor changing neutral currents (FCNC) can be suppressed by imposing $\mathcal{Z}_2$ symmetry on the Lagrangian.
That is the invariance on the Lagrangian under the interchange of $\Phi_1 \rightarrow \Phi_1$ and $\Phi_2 \rightarrow -\Phi_2$.
If it is allowed to violate the discrete $\mathcal{Z}_2$ symmetry softly, then FCNC are naturally suppressed at the tree level.
The $m^2_{12}$ term in Eq. \ref{eq:eq2} ensures the breaking of the discrete symmetry.
Therefore, if the discrete $\mathcal{Z}_2$ symmetry is extended to the Yukawa sector, we end up with four independent THDMs.
In a result, this discrete symmetry might be extended to the Higgs-fermion Yukawa interactions in a couple of different and independent ways.
In Type I, $\Phi_1$ couples to the all fermions and generates mass.
Type II is defined as where $\Phi_2$ couples to up-type quarks and $\Phi_1$ couples to down-type quarks and leptons \cite{Barger:1989fj,Aoki:2009ha,Branco:2011iw}.
In Type III, $\Phi_2$ couples to up-type quarks and to leptons and $\Phi_1$ couples to down-type quarks.
Lastly in type IV, $\Phi_2$ couples to all quarks and $\Phi_1$ couples to leptons.

In this work, we donÕt need to consider 14 parameters in the scalar potential. 
Since CP-violation doesn't have effect on the charged Higgs production at the tree level in this work but the neutral Higgses, it is not taken in to account.
Thus, $\lambda_6=\lambda_7$ are set to zero and the complex parameters $m_{12}$ and $\lambda_5$ are taken as real.
In a result, under these assumptions, the free parameter number reduces to 8.
According to the electroweak symmetry breaking, 3 degrees of freedom are eaten by the Goldstone bosons and give mass to the electroweak messenger particles.
The rest of the degree of freedom makes the prominent property of the model; two charged Higgs particles and 3 neutral Higgs particles \cite{Gunion:1989we}.
The 8 free parameters in the Higgs potential can be rewritten by eight physical mass parameters which are the masses of the neutral Higgs bosons $(m_{h/H^0/A^0/H^\pm})$ and charged Higgs bosons ($m_{H^\pm}$), the vacuum expectation value ($\nu=246 \; GeV$), the ratio of the vacuum expectation values ($\beta$), mixing angle between the CP-even neutral Higgs states ($\alpha$) and the soft breaking scale of the discrete symmetry $M^2=m_{12}^2/(\sin\beta\cos\beta)$ \cite{Gunion:2002zf}.
The free parameters introduced above in the THDM potential could be expressed in different basis, since the physics is not changed, with the help of \texttt{2HDMC} \cite{Eriksson:2009ws} we could convert these interchangeably.

\section{The Calculation of The Leading Order Cross Section of the $\gamma\gamma\rightarrow H^+H^-$}
\label{sec:3}

\begin{figure}[htbp]
\centering
	\includegraphics[width=0.60\textwidth]{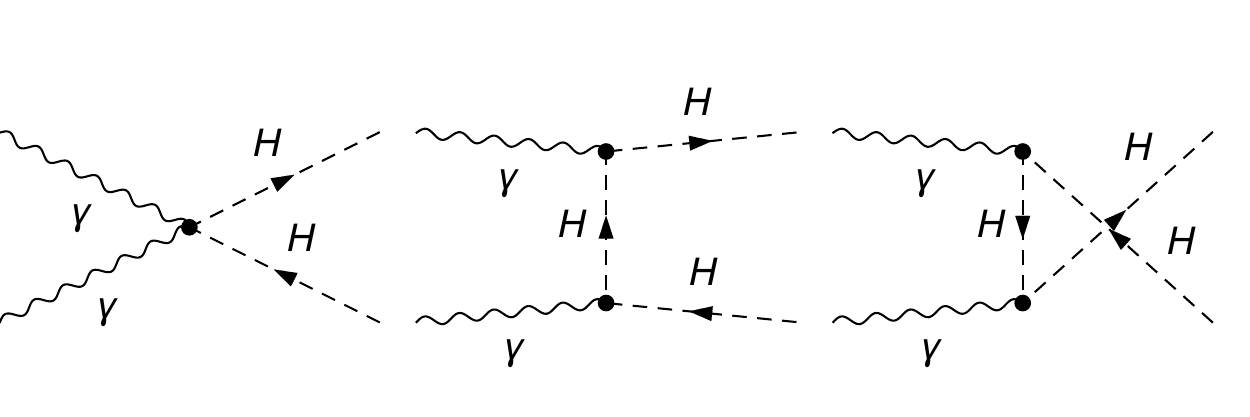}
\caption{Feynman diagrams contribution to the process $\gamma\gamma\rightarrow H^+H^-$ at the tree level.}
\label{fig:fig1}
\end{figure}

In this section, analytical expressions of the cross section in $e^+e^-$ collider for the charged Higgs pair production are given.
Throughout this paper, the process for the charged Higgs pair production is denoted as 
\begin{eqnarray}
\gamma (k_1,\mu)\; \gamma (k_2,\nu)\;\rightarrow \;  H^+ (k_3)\; H^- (k_4) \;\;\; (i,j=1,2)
\label{eq:eq3}
\end{eqnarray}
where $k_a$ $(a=1,...,4)$ are the four momenta of the incoming photons and outgoing charged Higgs bosons, respectively.
All the relevant Feynman diagrams contributing to the process $\gamma\gamma \rightarrow H^+\; H^-$ at the tree level are depicted in Fig \ref{fig:fig1}, where they are generated using \texttt{FeynArts}.
The expression of the amplitudes for each type of diagrams are constructed using \texttt{FeynArts}\cite{Kublbeck:1992mt,Hahn:2000kx}, the relevant part of the Lagrangian and the corresponding Feynman rules for the vertices are defined in Ref. \cite{Haber:1984rc}.
After the construction of the amplitudes, the simplification of the fermion chains and squaring the corresponding amplitudes are done using \texttt{FormCalc}\cite{Hahn:2006qw}.
Due to the photon coupling, there are three topological different diagrams at the tree level.
The total Feynman amplitude at this level is given below;
\begin{equation}
\mathcal{M}^0=\mathcal{M}_{\hat{q}}^0+\mathcal{M}_{\hat{t}}^0+\mathcal{M}_{\hat{u}}^0
\end{equation}
where $\mathcal{M}_{\hat{q}}^0$, $\mathcal{M}_{\hat{t}}^0$ and $\mathcal{M}_{\hat{u}}^0$ represents the amplitudes calculated from the quartic coupling, t-channel and u-channel Feynman diagram, respectively.
The explicit expressions for each channel in Fig. \ref{fig:fig1} are given as
\begin{eqnarray}
\mathcal{M}^0_{\hat q}&=&2 i e^2 g^{\mu\nu} \epsilon _\mu (k_1)\epsilon_\nu(k_2)\\
\mathcal{M}^0_{\hat t}&=&\frac{+ie^2}{\hat{t}-m_{H^+}^2}(k_1-2k_4)^\nu\epsilon_\nu(k_2)(k_2+k_3-k_4)^\mu \epsilon_\mu(k_1)\\ 
\mathcal{M}^0_{\hat u}&=&\frac{-ie^2}{\hat{u}-m_{H^+}^2}(k_1-2k_4)^\mu\epsilon_\mu(k_1)(k_1+k_3-k_4)^\nu \epsilon_\nu(k_2)
\end{eqnarray}
where $\hat{t}=(k_1-k_3)^2$ and $\hat{u}=(k_2-k_4)^2$ represents the Mandelstam variables.
After squaring the total amplitude and summing over the polarization vectors, the calculated expression is written as
\begin{eqnarray}
\left | \mathcal{M}^0\right |^2=&256 \pi ^2 \alpha^2 \left[ \left( \frac{t}{(t-m_{H^+}^2)^2} + \frac{u}{(u-m_{H^+}^2)^2} -  \frac{1}{t-m_{H^+}^2} - \frac{1}{u-m_{H^+}^2} \right) m_{H^+}^2  \right. \\ \nonumber
&\left. +\frac{ 9 m_{H^+}^4 -3 m_{H^+}^2 (2 m_{H^+}^2+s)+(s+t) (s+u)}{2 (t-m_{H^+}^2) (u-m_{H^+}^2)}\right]
 \end{eqnarray}

For the numerical calculation, the scattering amplitude is evaluated in the center of mass frame where the four-momentum and scattering angle are denoted by ($k, \theta$).
The energy $(k_i^0)$ and momentum $(\vec{k}_i)$ of the incoming and outgoing particles are given below in terms of CM energy :
\begin{eqnarray} \label{eq:cm}
& k_1=\frac{\sqrt{ s}}{2}(1,0,0,1),\;\;\;k_2=\frac{\sqrt{ s}}{2}(1,0,0,-1),\\
& k_3=(k_3^0,|\vec{k}| \sin\theta,0,|\vec{k}| \cos\theta),\\
& k_4=(k_4^0,-|\vec{k}| \sin\theta,0,-|\vec{k}|\cos\theta)\\
& k_3^0=\frac{ s+m_i^2-m_j^2}{2 \sqrt{ s}},~k_4^0=\frac{ s+m_j^2-m_i^2}{2 \sqrt{ s}},\\
& |\vec{k}|=\frac{\lambda( {s},m_{H^+}^2, m_{H^-}^2 ) }{\sqrt{ s}}.
\end{eqnarray}
where $m_i$ is the mass of the corresponding particle. The cross section of the process is calculated by taking into account the flux factor of the incoming particles and integral over the phase space of the outgoing particles as
\begin{equation}
{\hat{\sigma}}_{\gamma\gamma \rightarrow H^+H^-}({s})=\frac{ \lambda( {s},m_{H^+}^2, m_{H^-}^2 )}{16 \pi {s}^2}  \sum_{pol}{|\mathcal{M}^0|^2}\,,
\label{eq:partcross}
\end{equation}
where 
\begin{equation}
\lambda( {s},m_{H^+}^2, m_{H^-}^2 )=\sqrt{ ({s}-m^2_{H^+}-m^2_{H^-})^2-4m^2_{H^+}m^2_{H^-} }/2
\end{equation}
is the K\"allen function for the phase space of the outgoing charged Higgs pairs.

The total integrated cross section of the process in a $e^+e^-$-collider could be calculated by convoluting the cross section with the photon luminosities created by the backscattering electrons.
\begin{equation}
\sigma(s)=\int_{x_{min}}^{x_{max}} \hat{\sigma}_{\gamma\gamma\rightarrow  H^+H^-}( \hat{s};\; \hat{s}=z^2s ) \frac{dL_{\gamma\gamma}}{dz}\;dz\,,
\label{eq:foldcross}
\end{equation}
where $s$ and $\hat{s}$ are the CM energy in $e^+e^-$ collisions and $\gamma\gamma$ subprocess, respectively. 
$x_{min}$ is the threshold energy for production of the charged Higgs pair and defined as $x_{min}=(m_{H^+}+m_{H^-})/\sqrt{s}$.
The maximum fraction of the photon energy is taken as $x_{max}=0.83$ \cite{Telnov:1989sd}. 
The distribution function of the photon luminosity is given by
\begin{equation}
\frac{dL_{\gamma\gamma}}{dz}=2z\int_{z^2/x_{max}}^{x_{max}}\frac{dx}{x}F_{\gamma/e}(x)F_{\gamma/e}\left(\frac{z^2}{x}\right)\,,
\end{equation}
where $F_{\gamma/e}(x)$ is the energy spectrum of the Compton back scattered photons from initial unpolarized electrons and it is defined as a function of fraction $x$ of the longitudinal momentum of the electron beam \cite{Telnov:1989sd}.

\section{Numerical Results and Discussion}
\label{sec:4}
In this section, the numerical results for the charged Higgs pair production via photon-photon collisions are presented.
SM parameters are taken from Ref. \cite{Eidelman:2004wy} and fine-structure constant is $\alpha(m_Z)=1/127.934$. 
Apart from the SM parameters, we also need to set a value for the free parameters introduced by addition of the new Higgs doublet.
In the light of the results from the charged Higgs search in the previous experiments, we employed \texttt{HiggsBounds} \cite{Bechtle:2013wla} and \texttt{HiggsSignals} \cite{Bechtle:2013xfa} to determine a parameter set for the numerical analysis.
This set of parameters given below is not excluded yet by any experimental constraint.
We define the free parameters in the physical mass basis;
The mass spectrum of the neutral Higgs bosons are set to $m_{h^0/A^0/H^0}=125/400/400\;GeV$, the charged Higgs mass is taken as a free parameter in the results.
As it is mentioned before, the CP violation is not included in the calculation so that the $\lambda_6$ and $\lambda_7$ are set to 0.
Finally, the soft symmetry breaking term is taken as $m_{12}^2 = 15600$ and $\sin(\beta-\alpha)= 1$.
\begin{figure}[htbp]
\centering
\subfloat[]{\label{fig:fig2a}
\includegraphics[height=5.75cm]{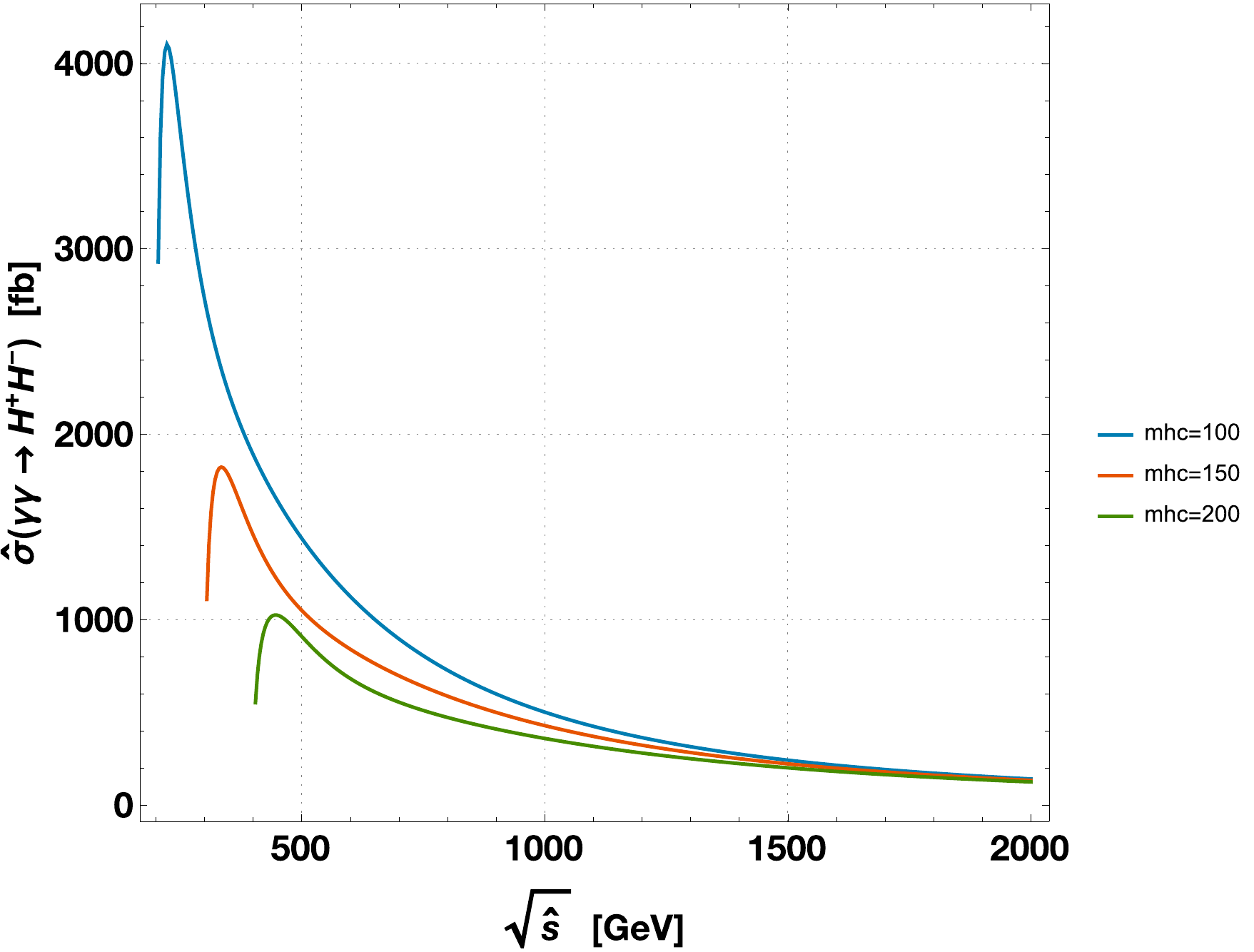}}
\subfloat[]{\label{fig:fig2b}
\includegraphics[height=5.75cm]{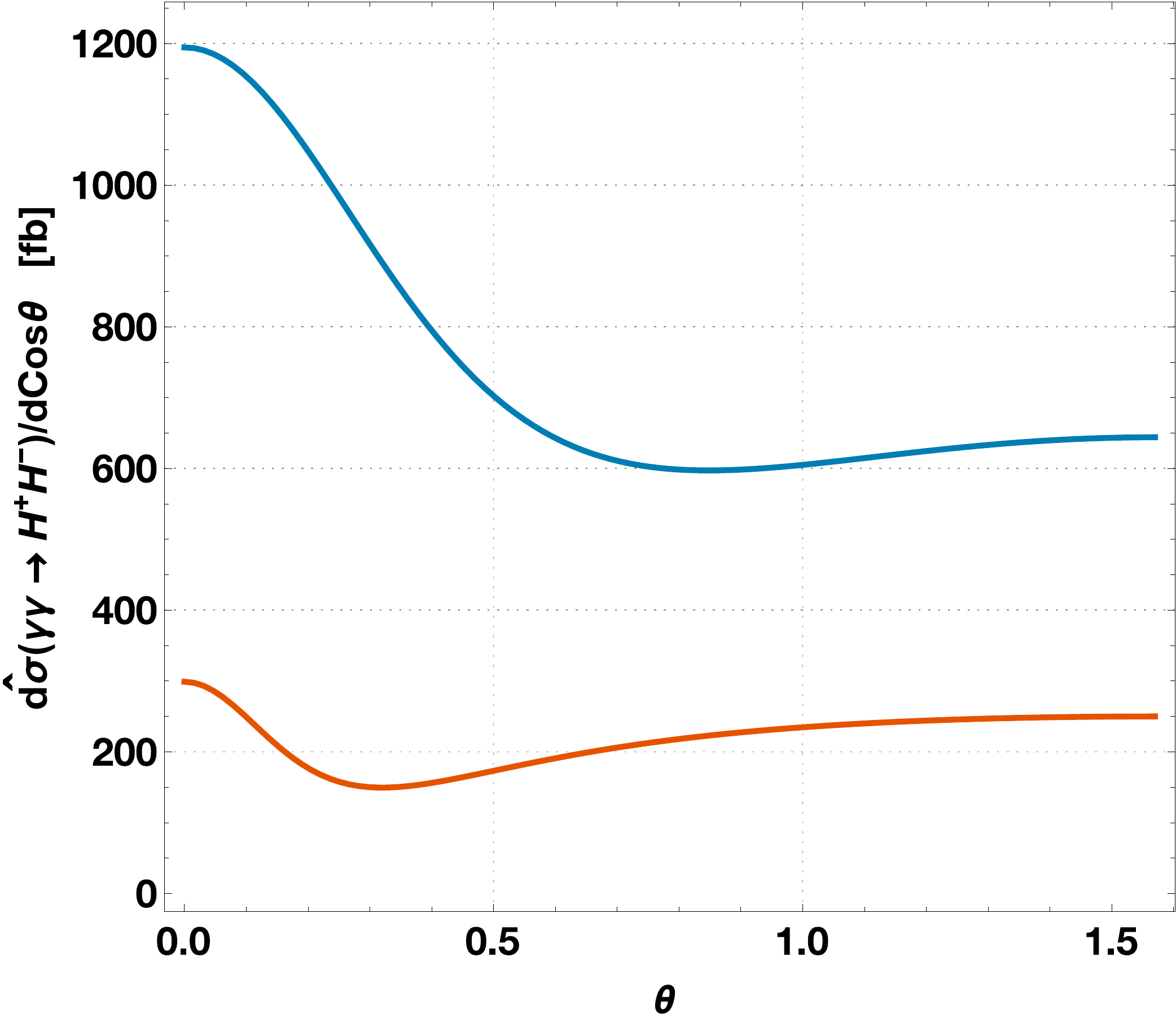}}
\caption{(a) Integrated cross section of the process $\gamma\gamma\rightarrow H^+H^-$ as a function of $\hat{s}$ for various charged Higgs masses. 
(b) Differential cross section as a function of the angle between the incoming photon and positive charged Higgs boson for $\sqrt{\hat{s}}=0.5\;TeV-1.0\;TeV$. 
Where the orange line stands for $\sqrt{\hat{s}}=0.5\;TeV$ and blue line stands for $\sqrt{\hat{s}}=1.0\;TeV$.}
\label{fig:fig2}
\end{figure}
After substituting all the parameters, the cross section of the process $\gamma\gamma \rightarrow H^+H^-$ is calculated and given as a function of CM energy in Fig. \ref{fig:fig2}(a) up to 2 TeV for three different charged Higgs masses.
As it can bee seen, the cross section declines gradually for higher charged Higgs mass.
The cross section have a peak around $\sqrt{\hat s}=220\; GeV$ and it gets value of $\hat{\sigma}\approx 4.15\; pb$ for $m_{H^\pm}=100\;{GeV}$, at high energies the cross section decays and it gets a value of $\hat{\sigma} (\sqrt{\hat {s}}=1\; TeV) \approx 0.5\; {pb}$.
The differential cross section as a function of angle between the incoming photon and positive charged Higgs boson for 0.5 TeV and 1 TeV CM energies are given in Fig. \ref{fig:fig2}(b) where the charged Higgs mass is set to $m_{H^\pm}=150\;{GeV}$.
In Fig. \ref{fig:fig2}(b), there is a small asymmetry at low CM energy, therefore the asymmetry gets large for higher CM energy due to the asymmetrical $t-$ and $u-$terms presented in the cross section. 
The differential cross section gets high at the tree level when $\theta\approx 0$.
It can be seen in Fig. \ref{fig:fig2}(b) that the production of the charged Higgs boson is produced in the forward and backward direction largely, and it will be much more possible to detect them in that region of the collision.

\begin{figure}[htbp]
\centering
	\includegraphics[height=6.0cm]{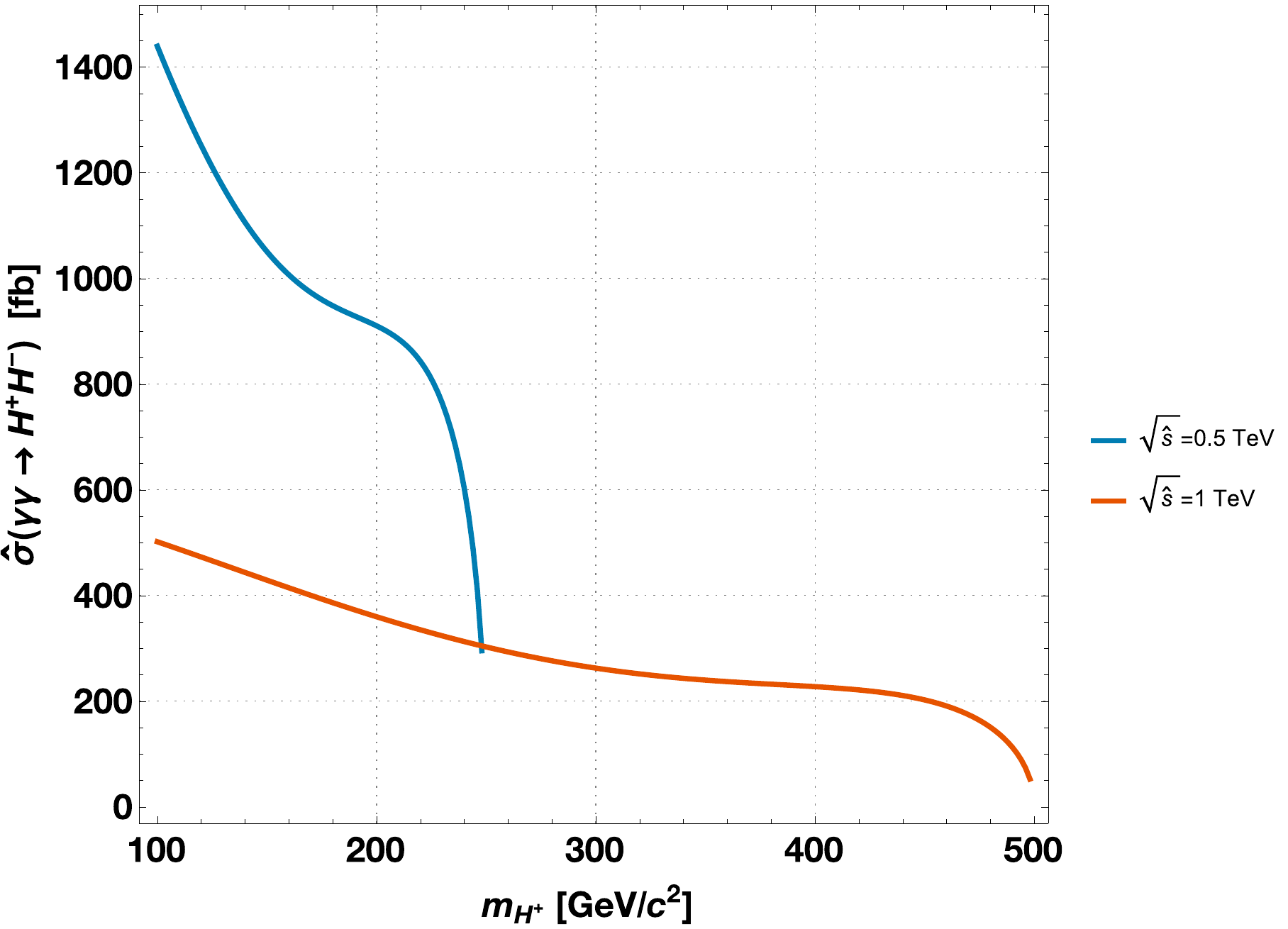}
\caption{
Integrated cross section of the process $\gamma\gamma\rightarrow H^+H^-$ as a function of $\tan\beta$ for $\hat{s}=0.5\;TeV-1.0\;TeV$. 
Where the orange line stands for $\hat{s}=0.5\;TeV$ and blue line stands for $\hat{s}=1.0\;TeV$.}
\label{fig:fig3}
\end{figure}
In Fig. \ref{fig:fig3}, the $m_{H^\pm}$ dependence of the cross section is plotted for $0.5\,{TeV}$ and $1\,{TeV}$ CM energies.
In each case, the cross section gets high for the low charged Higgs masses.
Therefore, the cross section is also high for the lower CM energy due to the $t-$ and $u-$channel Feynman diagrams.

\begin{figure}[htbp]
\centering
	\includegraphics[height=7.0cm]{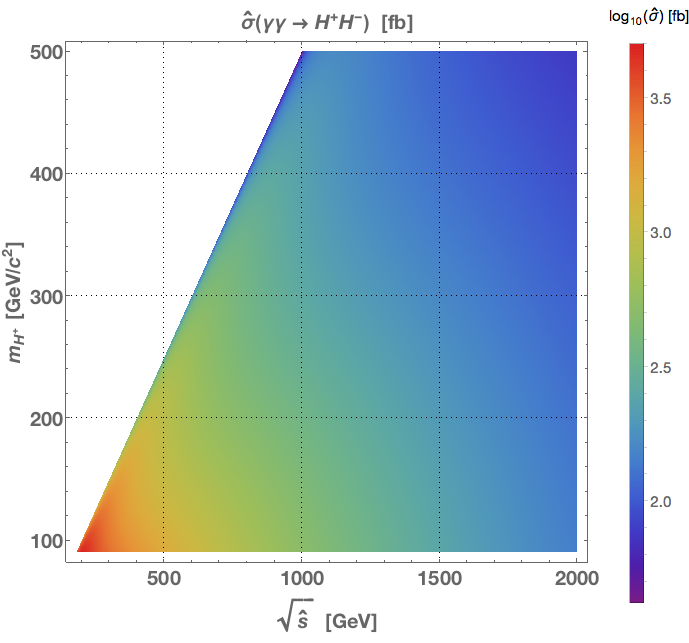}
\caption{
Integrated cross section of the process $\gamma\gamma\rightarrow H^+H^-$ as a function of $\hat{s}$ and $m_{H^\pm}$. 
}
\label{fig:fig4}
\end{figure}
Two dimensional analysis over the integrated cross section of the $\gamma\gamma \rightarrow H^+H^-$ process is drawn in Fig. \ref{fig:fig4} as a function of CM energy and the charged Higgs mass $m_{H^\pm}$.
Left up corner of the plot shows a region in white where the process is kinematically not accessible.
At the red region in Fig. \ref{fig:fig4}, the cross section gets a value up to $3.75\;{pb}$ at low charged Higgs masses and also at low CM energy.
Moving either on high CM and higher charged Higgs mass makes the cross section decline, dramatically.

\begin{figure}[htbp]
\centering
	\includegraphics[height=6.5cm]{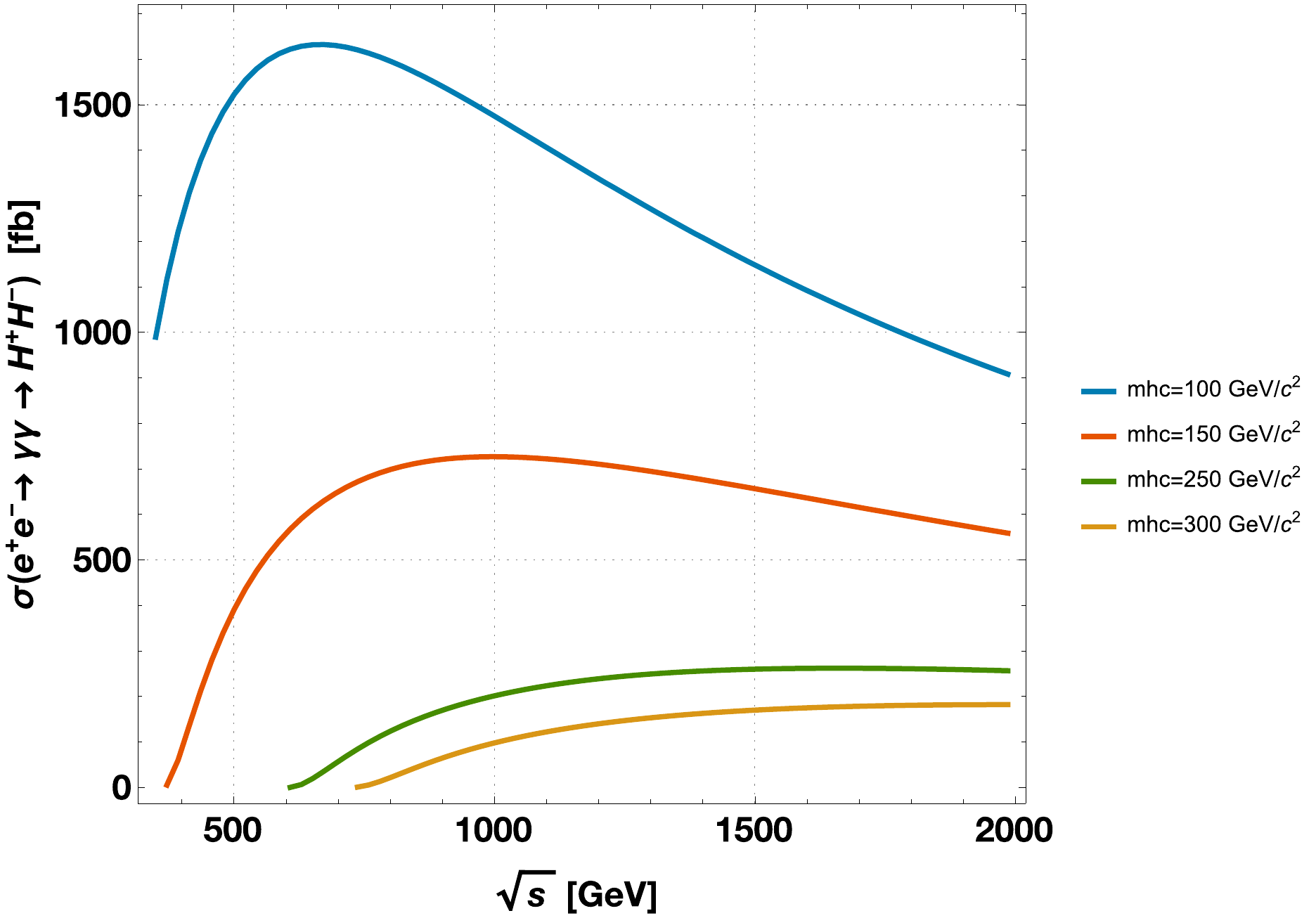}
\caption{Total integrated cross section of the process $e^+e^-\gamma\gamma\rightarrow H^+H^-$ as a function of CM energy and for various charged Higgs masses depicted by different color.}
\label{fig:fig5}
\end{figure}
Total integrated cross section of the process $\gamma\gamma\rightarrow H^+ H^-$ is also calculated where the process $\gamma\gamma\rightarrow H^+ H^-$ is taken as a subprocess in a $e^+e^-$-collider.
The cross section is convoluted with the photon luminosity of the back-scattered photons with the help of Eq. \ref{eq:foldcross}.
In Fig. \ref{fig:fig5}, the total integrated cross section is pictured up to $\sqrt{s}=2\;TeV$ for four different charged Higgs masses. 
As it is expected from the previous results, the total integrated cross section gets higher for the lower $m_{H^\pm}$.
The production rates around $\sqrt{s}=550\;{GeV}$ is $1.65\;{pb}$ and it declines down to $0.9 \;{pb}$ at $\sqrt{s}=2\;{TeV}$.

\section{CONCLUSION}
\label{sec:5}

LHC experiment affirmed the existence of a neutral Higgs boson \cite{Aad:2012tfa, Chatrchyan:2012ufa}, hereafter, the discovery of a charged Higgs boson or any charged Higgs like particle at the future colliders will be a clear evidence of the existence of new physics beyond the SM.
The resent announcement on excess in the two photon channel \cite{ATLAS:2015dxe, CMS:2015dxe} at ATLAS and CMS experiment could be a clear evidence on the CP-even or CP-odd Higgs boson predicted by the THDM.
Therefore, their masses have no impact on the charged Higgs pair production.

In this paper, we have calculated the charged Higgs boson pair production at the photon-photon colliders at the tree level in detail.
According to the picture drawn by the previous section, the charged Higgs boson pair production via photon-photon collisions have a nice production rates which can easily be detected and the properties could be studied thoroughly with the planned luminosity. 
The total integrated production rate of the process $e^+e^-\rightarrow\gamma\gamma\rightarrow H^+H^-$ reaches highest 1.63 pb at $\sqrt{s}=0.65$ TeV for the $m_{H^\pm}=100 { GeV}/c^2$ and falls down slowly at higher CM energies.

Given the parameters studied for the numerical analysis, the charged Higgs boson have a branching ratio close to 1 for the decay to $\tau$ and neutrino.
According to the free parameters in the THDM and in Type II, charged Higgs boson follows the decay chain $H^+\rightarrow \tau\nu_\tau$.
The signature of the charged Higgs boson pair production in the detector will be tagging two taus and missing energy related with the neutrinos.
Taking into account the acceptance of the detector, it could be a clear signal for charged Higgs boson as well as new physics beyond the Standard Model.
In a conclusion, a photon collider with the parameter set discussed above is a viable option for the $e^+e^-$-collider which could explore the low charged Higgs mass of the parameter space.
However, even if the ILC experiment could detect a charged Higgs boson, it still needs to be affirmed which model it belongs to, since many extensions of the SM like Supersymmetry \cite{Martin:1997ns} does predict a charged Higgs boson.


\section*{Acknowledgement}
Author thanks to Faculty of Science in Ege University for having access to the computing resource of \texttt{fencluster}.


\begin{thebibliography}{5}
\small
\bibitem{Branco:2011iw} 
  G.~C.~Branco, P.~M.~Ferreira, L.~Lavoura, M.~N.~Rebelo, M.~Sher and J.~P.~Silva,
  Phys.\ Rept.\  {\bf 516}, 1 (2012)
  doi:10.1016/j.physrep.2012.02.002
  [arXiv:1106.0034 [hep-ph]].
  
 \bibitem{Gunion:1989we} 
  J.~F.~Gunion, H.~E.~Haber, G.~L.~Kane and S.~Dawson,
  Front.\ Phys.\  {\bf 80}, 1 (2000).
  
  \bibitem{Abbiendi:2013hk} 
  G.~Abbiendi {\it et al.} [ALEPH and DELPHI and L3 and OPAL and LEP Collaborations],
  Eur.\ Phys.\ J.\ C {\bf 73}, 2463 (2013)
  doi:10.1140/epjc/s10052-013-2463-1
  [arXiv:1301.6065 [hep-ex]].

\bibitem{Abazov:2008rn} 
  V.~M.~Abazov {\it et al.} [D0 Collaboration],
  Phys.\ Rev.\ Lett.\  {\bf 102}, 191802 (2009)
  doi:10.1103/PhysRevLett.102.191802
  [arXiv:0807.0859 [hep-ex]].

\bibitem{Abazov:2009wy} 
  V.~M.~Abazov {\it et al.} [D0 Collaboration],
  Phys.\ Rev.\ D {\bf 80}, 051107 (2009)
  doi:10.1103/PhysRevD.80.051107
  [arXiv:0906.5326 [hep-ex]].

\bibitem{Abazov:2009ae} 
  V.~M.~Abazov {\it et al.} [D0 Collaboration],
  Phys.\ Rev.\ D {\bf 80}, 071102 (2009)
  doi:10.1103/PhysRevD.80.071102
  [arXiv:0903.5525 [hep-ex]].


\bibitem{Aaltonen:2009ke} 
  T.~Aaltonen {\it et al.} [CDF Collaboration],
  Phys.\ Rev.\ Lett.\  {\bf 103}, 101803 (2009)
  doi:10.1103/PhysRevLett.103.101803
  [arXiv:0907.1269 [hep-ex]].
  

\bibitem{Aad:2012tj} 
  G.~Aad {\it et al.} [ATLAS Collaboration],
  JHEP {\bf 1206}, 039 (2012)
  doi:10.1007/JHEP06(2012)039
  [arXiv:1204.2760 [hep-ex]].

\bibitem{Chatrchyan:2012vca} 
  S.~Chatrchyan {\it et al.} [CMS Collaboration],
  JHEP {\bf 1207}, 143 (2012)
  doi:10.1007/JHEP07(2012)143
  [arXiv:1205.5736 [hep-ex]].


\bibitem{Behnke:2013xla}
  T.~Behnke, J.~E.~Brau, B.~Foster, J.~Fuster, M.~Harrison, J.~M.~Paterson, M.~Peskin and M.~Stanitzki {\it et al.},
  arXiv:1306.6327 [physics.acc-ph].
  
\bibitem{Behnke:2013lya}
  T.~Behnke, J.~E.~Brau, P.~N.~Burrows, J.~Fuster, M.~Peskin, M.~Stanitzki, Y.~Sugimoto and S.~Yamada {\it et al.},
  arXiv:1306.6329 [physics.ins-det].

\bibitem{Hespel:2014sla} 
  B.~Hespel, D.~Lopez-Val and E.~Vryonidou,
  JHEP {\bf 1409}, 124 (2014)
  doi:10.1007/JHEP09(2014)124
  [arXiv:1407.0281 [hep-ph]].

\bibitem{Jiang:1997cg} 
  Y.~Jiang, L.~Han, W.~G.~Ma, Z.~H.~Yu and M.~Han,
  J.\ Phys.\ G {\bf 23}, 1151 (1997)
  [J.\ Phys.\ G {\bf 23}, 385 (1997)]
  doi:10.1088/0954-3899/23/4/001
  [hep-ph/9703275].


\bibitem{Hashemi:2013sja} 
  M.~Hashemi,
  Commun.\ Theor.\ Phys.\  {\bf 61}, no. 1, 69 (2014)
  doi:10.1088/0253-6102/61/1/11
  [arXiv:1310.7098 [hep-ph]].

\bibitem{Ma:1996nq} 
  W.~G.~Ma, C.~S.~Li and L.~Han,
  Phys.\ Rev.\ D {\bf 53}, 1304 (1996)
  [Phys.\ Rev.\ D {\bf 54}, 5904 (1996)]
  [Phys.\ Rev.\ D {\bf 56}, 4420 (1997)].
  doi:10.1103/PhysRevD.53.1304, 10.1103/PhysRevD.54.5904, 10.1103/PhysRevD.56.4420

\bibitem{BowserChao:1993ji} 
  D.~Bowser-Chao, K.~m.~Cheung and S.~D.~Thomas,
  Phys.\ Lett.\ B {\bf 315}, 399 (1993)
  doi:10.1016/0370-2693(93)91631-V
  [hep-ph/9304290].


\bibitem{Craig:2013hca} 
  N.~Craig, J.~Galloway and S.~Thomas,
  arXiv:1305.2424 [hep-ph].


\bibitem{Barger:1989fj}
  V.~D.~Barger, J.~L.~Hewett and R.~J.~N.~Phillips,
  Phys.\ Rev.\  D {\bf 41}, 3421 (1990)

\bibitem{Aoki:2009ha}
Mayumi Aoki, Shinya Kanemura, Koji Tsumura, Kei Yagyu,
Phys. Rev. \textbf{D80}, 015017 (2009),
arXiv:0902.4665 [hep-ph].

\bibitem{Gunion:2002zf} 
  J.~F.~Gunion and H.~E.~Haber,
  Phys.\ Rev.\ D {\bf 67}, 075019 (2003)
  doi:10.1103/PhysRevD.67.075019
  [hep-ph/0207010].
  
\bibitem{Eriksson:2009ws} 
  D.~Eriksson, J.~Rathsman and O.~Stal,
  Comput.\ Phys.\ Commun.\  {\bf 181}, 189 (2010)
  doi:10.1016/j.cpc.2009.09.011
  [arXiv:0902.0851 [hep-ph]].



\bibitem{Kublbeck:1992mt}
  J.~Kublbeck, H.~Eck and R.~Mertig,
  Nucl.\ Phys.\ Proc.\ Suppl.\  {\bf 29A} (1992) 204.
  
\bibitem{Hahn:2000kx}
  T.~Hahn,
  Comput.\ Phys.\ Commun.\  {\bf 140} (2001) 418
  [hep-ph/0012260].

     \bibitem{Haber:1984rc}
  H.~E.~Haber and G.~L.~Kane,
  Phys.\ Rept.\  {\bf 117} (1985) 75.
  
\bibitem{Hahn:2006qw}
  T.~Hahn and M.~Rauch,
  Nucl.\ Phys.\ Proc.\ Suppl.\  {\bf 157} (2006) 236
  [hep-ph/0601248].
  


 
  
 
\bibitem{Telnov:1989sd}
  V.~I.~Telnov,
  Nucl.\ Instrum.\ Meth.\ A {\bf 294} (1990) 72.

\bibitem{Eidelman:2004wy} 
  S.~Eidelman {\it et al.}  [Particle Data Group Collaboration],
  Phys.\ Lett.\ B {\bf 592}, 1 (2004).

\bibitem{Bechtle:2013wla} 
  P.~Bechtle, O.~Brein, S.~Heinemeyer, O.~Stål, T.~Stefaniak, G.~Weiglein and K.~E.~Williams,
  Eur.\ Phys.\ J.\ C {\bf 74}, no. 3, 2693 (2014)
  doi:10.1140/epjc/s10052-013-2693-2
  [arXiv:1311.0055 [hep-ph]].

\bibitem{Bechtle:2013xfa} 
  P.~Bechtle, S.~Heinemeyer, O.~Stål, T.~Stefaniak and G.~Weiglein,
  Eur.\ Phys.\ J.\ C {\bf 74}, no. 2, 2711 (2014)
  doi:10.1140/epjc/s10052-013-2711-4
  [arXiv:1305.1933 [hep-ph]].

\bibitem{Aad:2012tfa}
  G.~Aad {\it et al.}  [ATLAS Collaboration],
  Phys.\ Lett.\ B {\bf 716} (2012) 1
  [arXiv:1207.7214 [hep-ex]].
  
\bibitem{Chatrchyan:2012ufa}
  S.~Chatrchyan {\it et al.}  [CMS Collaboration],
  Phys.\ Lett.\ B {\bf 716} (2012) 30
  [arXiv:1207.7235 [hep-ex]].
  
  
\bibitem{ATLAS:2015dxe} 
  ATLAS Collaboration [ATLAS Collaboration],
  ATLAS-CONF-2015-081.

\bibitem{CMS:2015dxe} 
  CMS Collaboration [CMS Collaboration],
  CMS-PAS-EXO-15-004.


\bibitem{Martin:1997ns}
  S.~P.~Martin,
  Adv.\ Ser.\ Direct.\ High Energy Phys.\  {\bf 21} (2010) 1
  [hep-ph/9709356].


\end{thebibliography}

\end{document}